\documentclass[sigconf]{acmart}
\AtBeginDocument{%
  }

\copyrightyear{2026}
\acmYear{2026}
\setcopyright{cc}
\setcctype{by}
\acmConference[ICSE-FoSE '26]{2026 IEEE/ACM 48th International Conference on Software Engineering: Future of Software Engineering }{April 12--18, 2026}{Rio de Janeiro, Brazil}
\acmBooktitle{2026 IEEE/ACM 48th International Conference on Software Engineering: Future of Software Engineering (ICSE-FoSE '26), April 12--18, 2026, Rio de Janeiro, Brazil}
\acmDOI{10.1145/3793657.3793891}
\acmISBN{979-8-4007-2479-4/2026/04}

\begin{document}

\title{The Future of Software Engineering Conferences: A New Zealand Perspective}

\author{Kelly Blincoe}
\email{k.blincoe@auckland.ac.nz}
\affiliation{%
  \institution{University of Auckland}
  \country{New Zealand}
}

\author{Sherlock A. Licorish}
\email{sherlock.licorish@otago.ac.nz}
\affiliation{
  \institution{University of Otago}
  \country{New Zealand}
}

\author{Judith Fuchs}
\email{judith.fuchs@canterbury.ac.nz}
\affiliation{
  \institution{University of Canterbury}
  \country{New Zealand}
}

\author{Amjed Tahir}
\email{A.Tahir@massey.ac.nz}
\affiliation{
  \institution{Massey University}
  \country{New Zealand}
}

\renewcommand{\shortauthors}{Blincoe et al.}

\begin{abstract}
  Software engineering (SE) conferences are vital for knowledge exchange and collaboration, yet can also involve significant barriers for researchers in geographically distant regions such as New Zealand. We identify barriers such as high travel costs, misaligned academic calendars, and limited representation, and propose strategies including hybrid participation, cost-conscious venues, and governance reforms. 
  We make recommendations to promote equitable global participation and strengthen the SE research community.
\end{abstract}



\maketitle

\section{Introduction}

New Zealand's geographic isolation limits informal networking opportunities and collaborations. This makes conference attendance even more important for New Zealand SE researchers, in order to facilitate discoveries and global partnerships. However, there are challenges, which we describe in this paper. These challenges include high financial costs, long travel times, and misaligned academic calendars. Additionally, limited representation in steering committees and the concentration of conferences in North America and Europe exacerbate accessibility issues. While this paper focuses on New Zealand, these challenges are common across many countries in the Oceania region and other parts of the world that are geographically distant from major SE conference hubs (e.g., countries in Central Africa). 

In this paper, we summarize the challenges and opportunities, including hybrid participation models, greater geographic diversity of conferences, flexible pricing, cost-conscious venue choices, and improved governance practices, making the case for more inclusive partnerships and an enduring SE conference landscape. 

\begin{figure}[b]
  \centering \includegraphics[width=.75\linewidth]{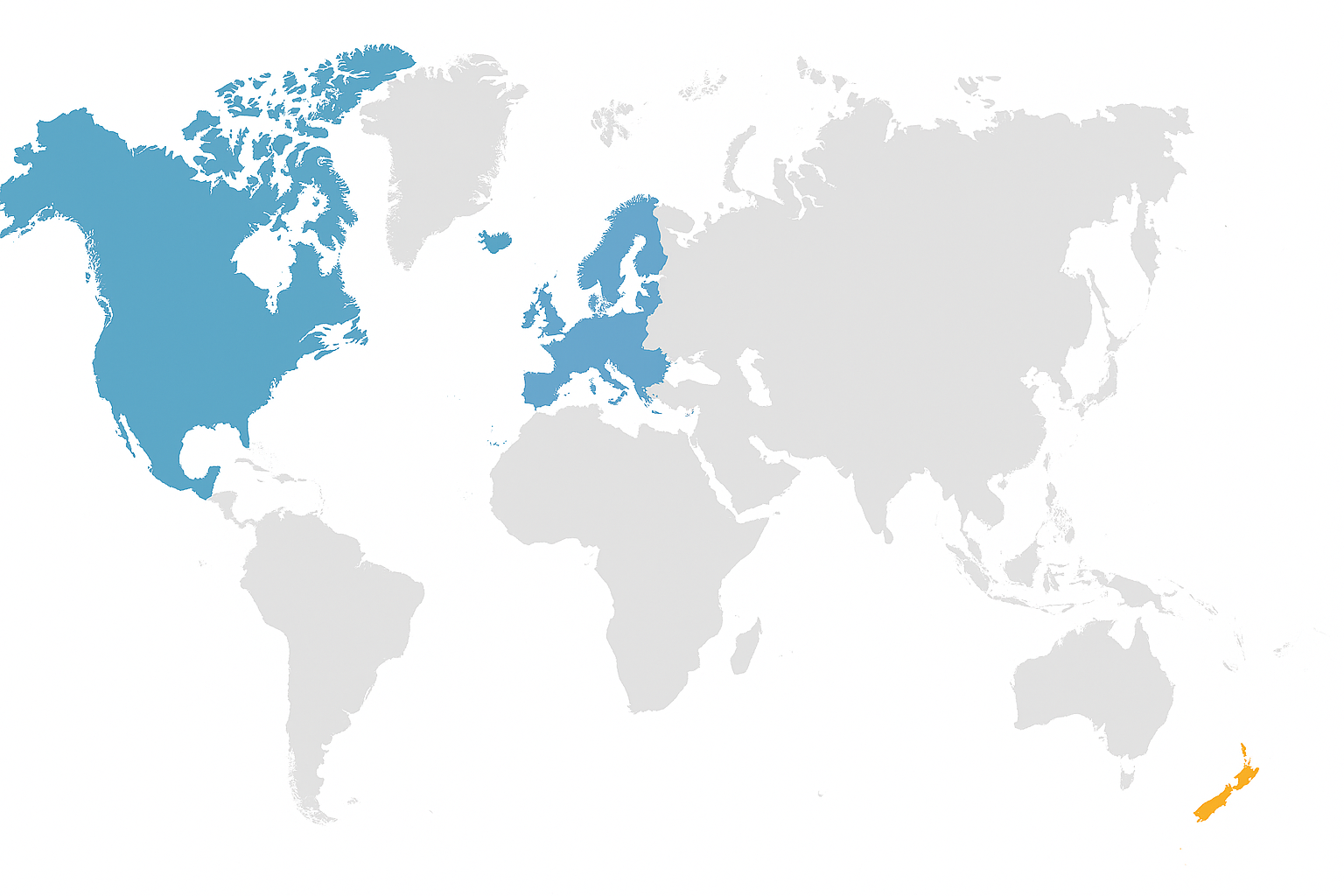}
  \caption{World map. North America and Europe in blue. The "Not North America/ Europe" region is shown in grey and orange (New Zealand). Generated by Microsoft Copilot.}
  \Description{World map highlighting regions of North America and Europe, illustrating large portion of world unhighlighted.}
  \label{world}
\end{figure}

\section{Challenges}

\subsection{Distance and travel challenges}
Many flagship SE conferences are held in North America or Europe. Since 2023, ICSE, FSE, and ASE operate on a rotating schedule ensuring ``that every region of the world has a major conference annually''.\footnote{https://sigsoft.medium.com/intentional-changes-underway-in-the-software-engineering-conferences-dedd6f8ba3c2} The rotation involves one of these conferences occurring in North America, one in Europe, and one in the "Not North America/Europe" region. While this is a good first step toward geographic diversity, the ``Not North America / Europe'' region represents a large majority of the world, as illustrated in Figure~\ref{world}. This means two-thirds of our flagship conferences are held in a rather small portion of the world. The vast size of the "Not North America/Europe" region, encompassing Oceania, Asia, South America, Africa, and the Middle East, means that many countries and many regions of the world still rarely host flagship conferences. Greater coordination among conferences could help achieve broader geographic diversity. This rotation has only been operating for four years including 2023-2026, resulting in four conferences being held in the ``Not North America / Europe'' region (one per year). Two of these were held in South America, both in Brazil, one in Asia (South Korea), and one in Oceania (Australia). Table~\ref{tab:topconf} documents locations of ICSE, FSE, and ASE from 2015-2026.

\begin{table}[ht]
    \centering
    \caption{Locations of ICSE, FSE, and ASE from 2015-2025}
    \label{tab:topconf}
    \begin{tabular}{p{0.15\linewidth} | p{0.75\linewidth}}\toprule
         \textbf{Conf.} & \textbf{Locations} \\ \midrule
    ICSE & Brazil (2026), Canada (2025, 2019), Portugal (2024), Australia (2023), USA (2022, 2016), Spain (2021), Korea (2020), Sweden (2018), Argentina (2017), Italy (2015) \\
    \hline
    FSE & Canada (2026), Norway (2025), Brazil (2024), USA (2023, 2020, 2018, 2016), Singapore (2022), Greece (2021), Estonia (2019), Germany (2017), Italy (2015) \\
    \hline
    ASE & Germany (2026), South Korea (2025), USA (2024, 2022, 2019, 2017, 2015), Luxembourg (2023), Australia (2021, 2020), France (2018), Singapore (2016) \\ \bottomrule
    \end{tabular}
\end{table}

 Beyond the financial implications of traveling long distances (which we discuss next), travel often involves flights exceeding 20 hours. The time commitment and jet lag associated with this travel reduce productivity and accessibility. Researchers with caregiving responsibilities or teaching obligations face additional barriers, often making in-person attendance impractical.

\subsection{Financial challenges}
Conference registration fees are typically in USD. When converted to NZD, these costs become disproportionately high. At the current exchange rate, for instance, attending ICSE 2026 and a two-day co-located conference as an ACM/IEEE member at early rates is \$1280USD, approximately \$2220NZD. 
Travel costs also quickly become prohibitive. Attending conferences from New Zealand involves flight costs often surpassing \$3,000NZD for trips to the U.S. and \$5,000NZD for trips to Europe.
This financial barrier is compounded by limited research funding compared to institutions in larger economies.\footnote{https://www.rnz.co.nz/news/business/489505/new-zealand-investment-in-research-and-development-half-of-other-leading-economies} 
For students and early career researchers (ECRs), those from smaller universities, and those without external research funding, these costs can be prohibitive, often resulting in missed opportunities for global engagement and partnerships. 

\subsection{Challenges to obtain a visa to participate}
North America and Europe have strict visa requirements for visitors. Visa requirements, costs, and short timelines can pose challenges, particularly for researchers without permanent residency or citizenship in their country of residence, which is often the case for many ECRs, including students. This may discourage or even prevent their attendance. The participation of ECRs could be encouraged by bringing conferences to countries that offer easy access in terms of travel requirements, issuing visa invitation letters substantially earlier to allow sufficient time to obtain visas, and providing sponsorship to cover visa costs.  

\subsection{Misaligned academic calendars}
New Zealand operates on a Southern Hemisphere academic calendar, with semesters typically running from late February to June and July to November. This schedule does not align with the Northern Hemisphere academic year, which most of the international SE research community follows. As a result, none of the flagship conferences are held during the Southern Hemisphere summer period (December to February) when New Zealand academics are less likely to have teaching obligations. Submission deadlines and reviewing timelines are also often less ideal. These misalignments often prevent full participation from New Zealand academics.  Multiple submission deadlines and the temporal distribution of conferences can alleviate some of these concerns.

\subsection{Participation in steering committees}
While some SE conferences have diverse steering committees (SCs) in terms of geography, there are many without any representation from Oceania and other regions. This means conferences will make decisions without considering the perspectives of SE researchers from all regions. Even when geographic diversity is ensured in SCs, time zone issues often prevent active participation if specific strategies are not employed. When North America and Europe's business hours are prioritized, it means members from Oceania, Asia, and other regions will have to participate in the middle of the night or not at all. 

\subsection{Skewed research participant demographics}
Conference papers often reflect the geographic context of their authors. As many submissions come from regions near the conference location, studies involving human participants may draw primarily from local populations, introducing an unintentional geographic bias. While this is not deliberate, it can influence research findings and the resulting tools and practices, highlighting the need for broader participation to ensure research outcomes reflect diverse global perspectives and worldviews. 

\section{Opportunities for the future}

\subsection{Hybrid/virtual conferences}
Virtual and hybrid conference models still tend to prioritize in-person attendees, leaving remote participants with fewer chances for meaningful engagement. Conferences should invest in high-quality virtual platforms that offer interactive sessions, networking lounges, and asynchronous engagement. This must be balanced with engaging in person events. This approach reduces cost and carbon footprint while maintaining global inclusivity. Regional conference models could also be considered, where satellite events are held across multiple regions. Frequent long-haul flights contribute significantly to carbon emissions, making hybrid and regional models not only inclusive but environmentally responsible.

\subsection{Enhance geographic diversity}
Another strategy can be to host conferences in diverse geographically dispersed locations. The geographic rotation program introduced in 2023 between ICSE, FSE, and ASE was a great first step. There is an opportunity to expand conference locations beyond North America and Europe to better reflect the global research community. Researchers from Asia, for example, represent a significant proportion of authors in our research community, but conferences are rarely held in this region. For New Zealand, most often only one or two researchers attend our flagship conferences in a given year. However, when ICSME 2025 was held in Auckland, 26 (13\%) participants came from New Zealand, illustrating that participation would increase if our flagship conferences were more accessible. These observations highlight the need for structural changes in how conferences are planned and hosted.

We acknowledge that limited mechanisms exist to facilitate greater geographic distribution, including limited capacity to compete with bids from larger countries and economies or researchers' lack of willingness to host large conferences in diverse locations. Conference leadership mechanisms could be re-envisioned, considering cases where General Chairs can be appointed independently of the host institution or region, and where hosting responsibilities are decoupled from local arrangements. This approach would allow conferences to be held in regions with fewer resources while still ensuring strong leadership and organizational quality. Additionally, global steering committees could proactively identify and support underrepresented regions through funding assistance, logistical support, and mentorship for local organizers. 

\subsection{Flexible pricing structures}
Conferences should consider differentiated registration fees based on geographic location and purchasing power parity to make conferences more accessible. Sponsorships and travel grants should prioritize researchers from underrepresented regions and lower-income countries, where financial constraints are often more severe. These measures would help ensure equitable participation across the global research community.

\subsection{Reduce overall costs}
Beyond geographic and currency considerations, conferences should rethink their cost structures. Many events are held in expensive hotels or convention centers, which significantly increase registration fees. Catered lunches and banquet dinners also add to costs.  Hosting conferences at universities or research institutes can reduce costs while maintaining professional standards. Building in longer lunch breaks where participants can source their own food from local eateries would reduce costs and allow participants to experience more authentic local cuisine. Such measures would also make conferences more accessible to participants from lower-income countries.

\subsection{Broadening participation and designing impactful conference tracks}
Securing funding for conference attendance often requires presenting accepted work and, in some cases, demonstrating strong potential for impact. Lowering barriers to presenting at conferences can broaden participation and foster diversity. One approach is to shift toward formats that emphasize discussion of ongoing work, with higher acceptance rates for short papers or posters, while reserving archival publications for journals. This model encourages meaningful dialogue and collaboration without excluding researchers who lack completed studies. In addition, conference tracks and workshops may be designed to encourage and deliver impact that is attractive to government, policy makers, and potential funders (including university leadership). For example, events that lead to tangible exchanges may help potential attendees to make a strong case and compete for limited funding (e.g., hackathons may be included that align with global priorities where tangible solutions/prototypes may be delivered). 

\subsection{Promote diversity in steering committees}
Conference organization and planning should actively involve a broad range of regional stakeholders in SC roles to ensure diverse perspectives are represented. This approach allows challenges to be identified and addressed early, reducing barriers and enhancing participation. Diversity should extend beyond geography to include career stages, ensuring that the needs of ECRs, who often face greater funding and teaching constraints, are considered. Current SCs frequently consist of senior researchers with strong funding histories, which may limit awareness of the broader community’s challenges. However, simply creating a diverse SC is not enough; mechanisms must enable meaningful participation. Time zone equity is critical; meetings should be scheduled to accommodate global members. A good example of this is the ICSE SC, which offers two time slots for each meeting. These practices help ensure that diversity is not only symbolic but functional, fostering inclusive decision-making and equitable representation.

\subsection{Supporting early career researchers}

SE conferences should encourage ECRs worldwide to take up leadership roles such as Program Committee (PC) memberships to support their professional development. The ICSE shadow PC and MSR Junior PC are good examples of such initiatives. ECRs could also be encouraged to take on leadership roles by organising networking events for other ECRs during conferences, performing some local organisation duties for conferences, and acting as session chairs. These initiatives could be expanded further to accommodate ECRs from geographically distant regions. For example, travel grants could support ECRs from distant locations to reduce some of the barriers to participating in conferences, thereby opening up opportunities for networking, organising, and session chairing. Furthermore, ECRs from geographically distant regions could be involved in promoting conferences online and also in leadership roles within virtual or hybrid conference models.

\section{Conclusion}
The future of SE conferences must prioritize inclusivity, sustainability, and equity. Addressing barriers such as geographic isolation, financial constraints, and academic calendar misalignment will benefit not only New Zealand but also many other regions outside North America and Europe. Implementing hybrid models, hosting conferences in diverse locations, reducing costs through university venues, and promoting regional representation in steering committees are practical steps toward this goal. Such changes can enable broader participation, reduce environmental impact, and ensure that the global SE community thrives through diverse perspectives and collaborations. 


\end{document}